\documentclass[aps,prl,twocolumn,showpacs,preprintnumbers,amsmath,amssymb]
{revtex4}
\usepackage{graphicx}
\usepackage{dcolumn}
\usepackage{bm}

\begin{document}
\title
{{\it Ab initio} calculation of intrinsic spin Hall conductivity of Pd and Au}
 
\author{G. Y. Guo}
\email{gyguo@phys.ntu.edu.tw}
\affiliation{Department of Physics and Center for Theoretical Sciences, 
National Taiwan University, Taipei 106, Taiwan}

\date{\today}   

\begin{abstract}
An {\it ab initio} relativistic band structure calculation of spin Hall conductivity
(SHC) ($\sigma_{xy}^z$) in Pd and Au metals has been performed. It is found that at low temperatures,
intrinsic SHCs for Pd and Au are, respectively, $\sim 1400 (\hbar/e)(\Omega {\rm cm})^{-1}$
and $\sim 400 (\hbar/e)(\Omega {\rm cm})^{-1}$. The large SHC in Pd comes from the 
resonant contribution from the spin-orbit splitting of the doubly degenerated 4$d$
bands near the Fermi level at symmetry $\Gamma$ and X points, and the smaller SHC in Au
is due to the broad free electron like 6$s$6$p$-bands. However, as the temperature
increases, the SHC in Pd decreases monotonically and reduces to 
$\sim 330 (\hbar/e)(\Omega {\rm cm})^{-1}$ at 300 K, while the SHC in Au increases steadily
and reaches $\sim 750 (\hbar/e)(\Omega {\rm cm})^{-1}$ at room temperature.
This indicates that the gigantic spin Hall effect 
[$\sigma_{xy}^z \approx 10^5 (\hbar/e)(\Omega {\rm cm})^{-1}$] 
observed recently in the Au/FePt system [T. Seki, {\it et al.}, Nature Materials {\bf 7}, 125 (2008)] 
is due to the extrinsic mechanisms such as the skew scattering by the impurities in Au.
\end{abstract}
\pacs{71.15.Rf, 72.15.Eb, 72.25.Ba}
\maketitle

Spin transport electronics (spintronics) has recently become a 
very active research field in condensed matter and materials physics
because of its potential applications in information storage and
processing and other electronic technologies~\cite{pri98} and also
because of many fundamental questions on the physics of electron
spin~\cite{zut04}. Spin current generation is an important issue
in the emerging spintronics. Recent proposals of the intrinsic
spin Hall effect (SHE) are therefore remarkable~\cite{mur03,sin04}. In
the SHE, a transverse spin current is generated in
response to an electric field in a metal with relativistic
electron interaction (spin-orbit coupling). This effect was first
considered to arise extrinsically, i.e., by impurity
scattering~\cite{dya71,hir99}. The scattering becomes spin-dependent in
the presence of spin-orbit coupling (SOC), and this gives rise to the SHE.
In the recent proposals, in contrast, the spin
Hall effect can arise intrinsically in hole-doped ($p$-type) bulk
semiconductors~\cite{mur03} and also in electron-doped ($n$-type)
semiconductor heterostructures~\cite{sin04} due to intrinsic
SOC in the band structure. The intrinsic SHE can be 
calculated and controlled,
whereas the extrinsic SHE depends sensitively on details of 
the impurity scattering. Therefore, the intrinsic SHE 
is more important for applicational purposes, 
in comparison with the extrinsic SHE.

In semiconductors, there have been experimental reports on the SHE
in $n$-type GaAs \cite{kat04}, $p$-type GaAs \cite{wun04} and
$n$-type InGaN/GaN superlattices \cite{cha07} in recent years.  
These experimental results have been 
discussed theoretically, and it is now recognized that the 
SHE in $n$-type GaAs is due to the extrinsic mechanisms, i.e., skew 
scattering and side-jump contributions \cite{ino04,Engel}, 
while that in $p$-type GaAs is mainly caused by the 
intrinsic mechanism \cite{mur04b,Onoda05}. 
This conclusion is supported by theoretical analysis 
of the impurity effect and vertex correction to 
SHE in the Rashba and Luttinger models \cite{ino04,mur04b}. In 
$p$-GaAs the fourfold degeneracy at 
the $\Gamma$-point of the valence bands acts as the Yang-monopole, which 
enhances the SU(2) non-Abelian Berry curvature \cite{mur04}.

On the other hand, the SHE in metallic systems is currently 
attracting interest, stimulated by latest experimental reports on 
the SHE or inverse spin Hall effect (ISHE),
i.e., the transverse voltage drop due to the spin current 
\cite{Saitoh06,Kimura07,Valenzuela06}.
The spin Hall conductivity (SHC) in metals is much larger than that in the 
semiconductors. Naively, this may be due to the large number of 
carriers, but the band structure is very important.\cite{guo08} 
Furthermore, the Fermi degeneracy temperature is much higher 
than room temperature, and hence the quantum coherence is more 
robust against the thermal agitations compared with the 
semiconductors systems. 

In recent experiments for metallic systems\cite{Saitoh06,Kimura07,Valenzuela06},
platinum shows the prominent SHE surviving even up to room 
temperature, whereas aluminum and copper show relatively
tiny SHE. However, the mechanism of the SHE in metals has 
been assumed to be extrinsic. In Ref.~\cite{Kimura07} this 
difference is attributed to a magnitude of spin-orbit coupling 
for each metal. However, platinum seems to be special even 
among heavy elements. This material dependence strongly suggests 
a crucial role of intrinsic contributions. Therefore it is 
highly desired to study the intrinsic SHE of platinum as a 
representative material for metallic SHE. If this analysis 
successfully explains the experiment, it will open up the possibility 
to theoretically design the SHE in metallic systems.
Therefore, we recently carried out {\it ab initio} calculations 
for the SHC in platinum.\cite{guo08} We found that the intrinsic SHC
is as large as $\sim 2000 (\hbar/e)(\Omega {\rm cm})^{-1}$
at low temperature, and decreases down to
$\sim 200 (\hbar/e)(\Omega {\rm cm})^{-1}$ at room temperature.
It is due to the resonant contribution from the spin-orbit splitting
of the doubly degenerated $d$-bands at high-symmetry
$L$ and $X$ points near the Fermi level. We showed, by modeling these 
near-degeneracies by an effective Hamiltonian, that the SHC 
has a peak near the Fermi energy and that the vertex correction due 
to impurity scattering vanishes, indicating that the large 
SHE observed experimentally in platinum\cite{Kimura07} is of intrinsic nature.

\begin{figure}[h]
\includegraphics[width=8cm]{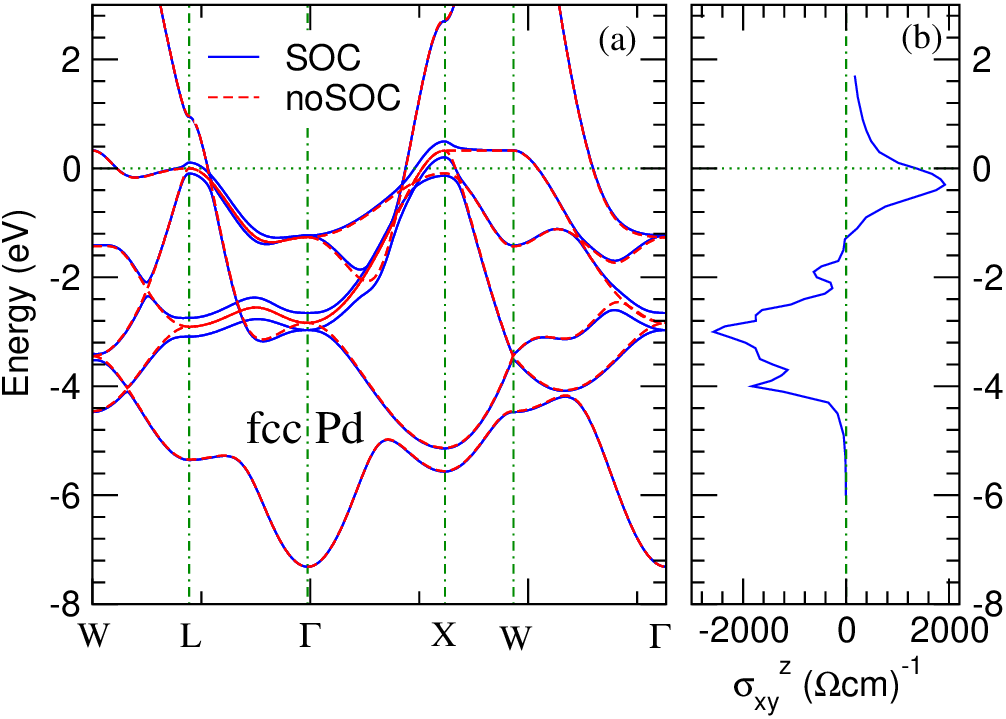}
\caption{\label{bs1} (color online) (a) 
 Relativistic band structure and (b) spin Hall conductivity 
 of fcc Pd. The zero energy and the dotted line
 is the Fermi level.
 The dashed curves in (a) are the scalar-relativistic
band structure.
}
\end{figure}

In this paper, we study the intrinsic SHE in Pd and Au metals
with first-principles band structure calculations.
The band structures of Pd and Au have been calculated using a 
fully relativistic extension~\cite{ebe88}
of the all-electron linear muffin-tin orbital
method~\cite{and75} based on the density functional theory with 
local density approximation~\cite{vos80}. 
The experimental lattice constants for Pd and Au used 
are 3.89 and 4.08 \AA, respectively.
The basis functions used are $s$, $p$, $d$ and $f$ 
muffin-tin orbitals~\cite{and75}.
In the self-consistent electronic structure calculations, 
89 $k$-points in the fcc irreducible wedge (IW)
of the BZ were used in the tetrahedron
BZ integration.  The SHC is evaluated by 
the Kubo formula, as described in Ref. \onlinecite{guo05}.
A fine mesh of  60288 $k$-points on a larger IW 
(three times the fcc IW) is used.  These numbers correspond to the  
division of the $\Gamma X$ line into 60 segments (see Fig.~\ref{bs1}).

Fig. 1 shows the relativistic band structure of Pd, and also the 
SHC ($\sigma_{xy}^z$) as a function of $E_F$. 
Clearly, the SHC peaks just below at the true Fermi level (0 eV), 
with a large value of $\sim$1400 $(\hbar /e)(\Omega{\rm cm})^{-1}$. 
This large value of the SHC is smaller than that of Pt~\cite{guo08}.
Nonetheless, it is still orders of magnitude larger
than the corresponding value in $p$-type semiconductors 
Si, Ge, GaAs and AlAs~\cite{guo05,yao05}. 
Note that the SHC in Pd decreases monotonically as 
the $E_F$ is artificially raised and becomes rather small above 3.0 eV. 
When the $E_F$ is artificially lowered, the SHC first peaks just below
the $E_F$ (-0.3 eV), then decreases considerably, and changes its
sign at $-$1.2 eV. As the $E_F$ is further lowered,
the SHC increases in magnitude again,
and becomes peaked at $-$3.0 eV with a large value of
$-$2600 $(\hbar /e)(\Omega{\rm cm})^{-1}$. The SHC
decreases again when the $E_F$ is further lowered,
and finally becomes very small below $-$5.0 eV.
Note that the bands below $-$5.5 eV and also above 0.5 eV
are predominantly of 5$s$ character and the effect of the
SOC is small. 

\begin{figure}[h]
\includegraphics[width=8cm]{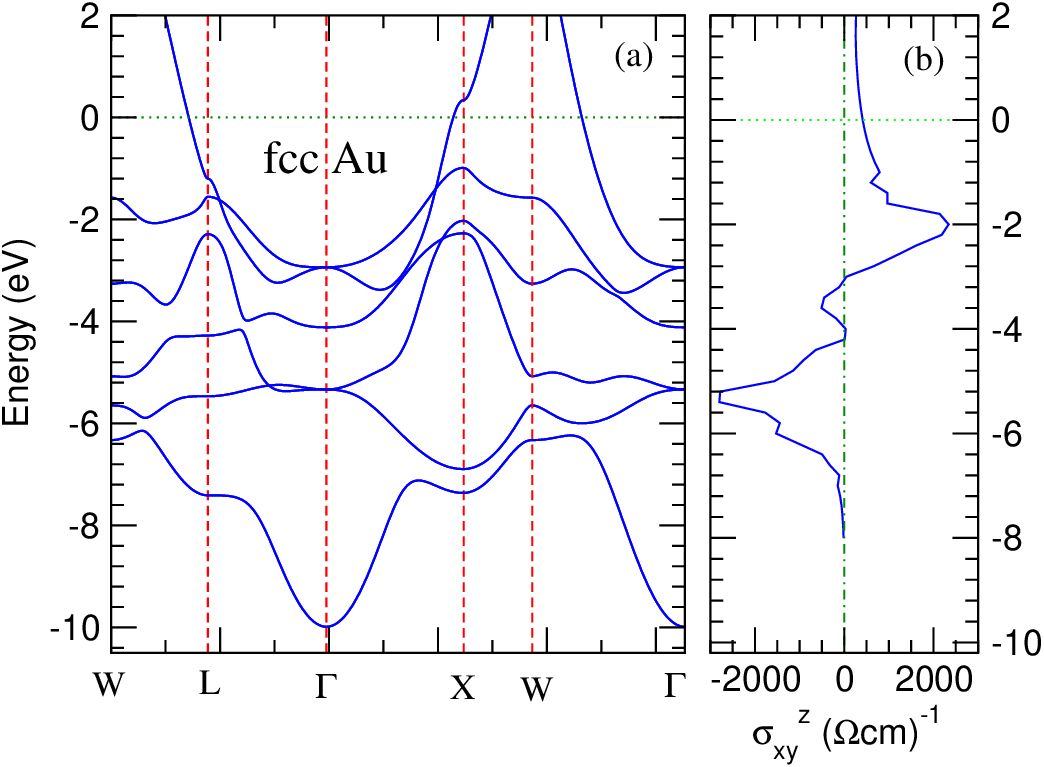}
\caption{\label{bs2} (color online) (a)
 Relativistic band structure and (b) spin Hall conductivity
 of fcc Au. The zero energy and the dotted line
 is the Fermi level.}
\end{figure}

We notice that a peak in the SHC appears at the double
degeneracies on the $L$ and $X$ points near 
$E_F$ in the scalar-relativisitc band structure 
(i.e., without the SOC) 
while the other peak at $-$3.0 eV occurs near the double
degeneracies at the $L$ and $\Gamma$ points (see Fig.~\ref{bs1}).
The double degeneracy (bands 5 and 6) at
$L$ is made mostly of $d_{x'z'}$ and $d_{y'z'}$ ($z'$: threefold 
axis), being consistent with the point group $D_{3d}$ at $L$. 
The double degeneracy (bands 4 and 5) at
$X$ consists mainly of $d_{x'z'}$ and $d_{y'z'}$ ($z'$: fourfold 
axis),
being consistent with the point group $D_{4h}$.
These double degeneracies are lifted by
the SOC, with a rather large spin-orbit splitting.
As in Pt, the large SHC in Pd may be attributed to these double 
degeneracies.\cite{guo08} 

The relativistic band structure of Au, and also the
SHC ($\sigma_{xy}^z$) as a function of $E_F$ are displayed in Fig. 2.
It is clear that both the shape and amplitude of the SHC versus 
$E_F$ curve (Fig. 2b) of Au are very similar to that of Pd (Fig. 1b)
and Pt~\cite{guo08}. This is because the band structure of Au (Fig. 2a)
is rather similar to that of Pd (Fig. 1a) and Pt\cite{guo08}. 
However, because Au has an extra valence electron and hence its
$d$-band is completely filled, the $E_F$ falls 
in the broad 6$s$6$p$ band where the SOC is relatively small.
As a result, the SHC in Au is relatively small at low temperatures
[$\sigma_{xy} (T=0$K$) \approx 400 (\hbar /e)(\Omega{\rm cm})^{-1}$],
being consistent with the previous {\it ab initio} calculation\cite{yao05}.
Nonetheless, it is a few times larger than the SHC in 
semiconductors~\cite{guo05,yao05}.

\begin{figure}[h]
\includegraphics[width=7cm]{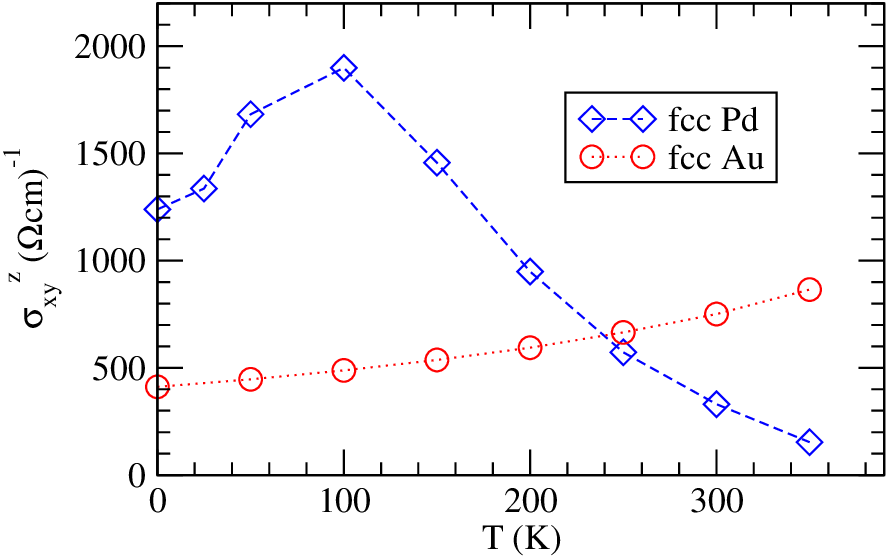}
\caption{\label{t-dependence}
(color online) Temperature-dependence of the
spin Hall conductivity $\sigma^{z}_{xy}$ in
Pd and Au metals.}
\end{figure}

The SHC can be written in terms of Berry curvature 
$\Omega_n^z({\bf k})$ as 
\begin{eqnarray}
\sigma_{xy}^{z} = \frac{e}{\hbar}\sum_{\bf k}\Omega^z({\bf k}) 
= \frac{e}{\hbar}\sum_{\bf k}\sum_n f_{{\bf k}n}\Omega_n^z({\bf k}),\nonumber \\
\Omega_n^z({\bf k}) = \sum_{n'\neq n}
\frac{2{\rm Im}[\langle{\bf k}n|j_x^z|{\bf k}n'\rangle\langle{\bf k}n'|v_y|{\bf k}n\rangle]}
 {(\epsilon_{{\bf k}n}-\epsilon_{{\bf k}n'})^2},
\end{eqnarray}
where the spin current operator $j_x^z = \frac{1}{2}\{s_z, {\bf v}\}$, 
with spin $s_z$ given by
$s_z=\frac{\hbar}{2}\beta\Sigma_z$ ($\beta$,  
$\Sigma_z$: $4\times 4$ Dirac matrices) 
\cite{guo05}.
$f_{{\bf k}n}$ is the Fermi distribution function for the $n$-th band at 
${\bf k}$. ${\Omega_n}^{z}$ can be regarded as
an analogue of the Berry curvature for the $n$-th band, and it is
enhanced when other bands come close in energy (i.e. near degeneracy). 
The SHC for Pd and Au calculated as a function of temperature
using Eq.(1) is shown in Fig. 3. 
Fig.~3 shows that the SHC in Pd decreases substantially as the temperature
($T$) is raised above 100 K, although it increases with temperature 
below 100 K. This rather strong temperature dependence is also 
due to the near degeneracies since the small energy scale is relevant
to the SHC there. Nevertheless, the SHC at room $T$
[$\sigma_{xy} (T=300$K$) = 350 (\hbar /e)(\Omega{\rm cm})^{-1}$] is
still rather large. Interestingly, in contrast, the SHC in Au increases
steadily with temperature to reach a value of 750 $(\hbar /e)(\Omega{\rm cm})^{-1}$.
This is because the $E_F$ cuts across the broad 6$s$6$p$ band in Au.
As a result, the intrinsic SHC in Au at room temperature is in fact larger than
that of Pd and Pt.

Excitingly, giant spin Hall effect at room temperature has been recently observed 
in a multi-terminal device with a Au Hall cross and an FePt perpendicular spin 
injector.\cite{sek08} The measured $\sigma^z_{xy} \approx 10^5(\hbar /e)(\Omega{\rm cm})^{-1}$ 
is orders of magnitude larger than the intrinsic SHC in bulk Au reported above.
This obviously indicates that the SHC due to intrinsic SOC in the band structure
of pure Au is not the dominant mechanism in the Au/FePt system.
Indeed, the authors of Ref. \onlinecite{sek08} attributed the giant SHE to
the extrinsic mechanism of the skew scattering by impurities in Au. Nonetheless,
its microscopic origin remains a puzzle and is currently 
under intensive investigation.\cite{naoto}

The author thanks N. Nagaosa, S. Murakami, T.-W. Chen and S. Maekawa
for stimulating discussions and collaboration on spin Hall effect in metals.
The author also thanks National Science Council of Taiwan for support, 
and also NCHC of Taiwan for the CPU time.

\end{document}